\documentclass[pre,showpacs,floatfix]{revtex4}
\usepackage{graphicx}

\def\be{\begin{equation}}
\def\ee{\end{equation}}
\def\bea{\begin{eqnarray}}
\def\eea{\end{eqnarray}}

\begin{document}

\title{Two-dimensional tunnel bifurcations with dissipation}

\author{A. K. Aringazin}
 \email{aringazin@mail.kz}
 \altaffiliation[Also at ]
 {Kazakhstan Division, Moscow State University, Moscow 119899,
Russia.}
 \affiliation{Department of Theoretical Physics, Institute
for Basic Research, Eurasian National University, Astana 473021,
Kazakhstan}
 \homepage{http://www.emu.kz}

 \author{Yuri Dahnovsky}
 \email{yurid@uwyo.edu}
 \affiliation{Department of Physics and Astronomy, P.O. Box 3905,
University of Wyoming, Laramie, WY 82071, USA}

 \author{V. D. Krevchik}
 \email{physics@diamond.stup.ac.ru}
 \affiliation{Department of Physics,
Penza State University, 40 Krasnaya St., Penza 440017, Russia}

 \author{M. B. Semenov}
 \email{physics@diamond.stup.ac.ru}
 \altaffiliation[Also at ]{Institute for Basic Research, P.O. Box
1577, Palm Harbor, FL 34682, USA.}
 \affiliation{Department of
Physics, Penza State University, 40 Krasnaya St., Penza 440017,
Russia}

 \author{V. A. Veremyev}
 \email{physics@diamond.stup.ac.ru}
 \affiliation{Department of Physics, Penza State University, 40
Krasnaya St., Penza 440017, Russia}

\author{A. A. Ovchinnikov}
 \affiliation{Joint Institute of Chemical Physics,
Kosygin Street 4, 117334 Moscow, Russia}

 \author{K. Yamamoto}
 \affiliation{Research Institute of International Medical Center of
Japan, Tokyo, Japan}

\date{August 20, 2003}

\begin{abstract}
Two-particle tunneling in synchronous and asynchronous regimes is
studied in the framework of dissipative quantum tunneling. We show
that the use of the proposed model is justified by a comparison
with realistic potential energy surfaces of porphyrin and
experimental dependence of the reaction rate on temperature. The
critical temperature $T_c$ corresponding to a bifurcation of the
underbarrier trajectory is determined. The effect of a heat bath
local mode on the probability of two-dimensional tunneling
transfer is also investigated. At certain values of the
parameters, the degeneracy of antiparallel tunneling trajectories
is important. Thus, four, six, twelve, etc., pairs of the
trajectories should be taken into account (a cascade of
bifurcations). For the parallel particle tunneling the bifurcation
resembles phase transition of a first kind, while for the
antiparallel transfer it behaves as second order phase transition.
The proposed theory allows for the explanation of experimental
data on quantum fluctuations in two-proton tunneling in porphyrins
near the critical temperature.
\end{abstract}


\maketitle

\section{\label{Sec1} Introduction}

Quantum tunneling dynamics of a particle interacting with a heat
bath is one of the important problems of modern condensed matter
physics
\cite{1}-\cite{6}. The interest to this problem is related to
studies in low-temperature superconductive tunnel junctions
\cite{10,11,14}, a dissipative quantum tunneling in crystals
\cite{18}, and low-temperature chemical reactions
\cite{1,3,4,5,23,24,dos,6}. In low dimensional systems an
effective mass approximation often fails. Thus, quantum tunneling
with dissipation becomes an important tool in the description of
electron transfer \cite{8,18}. In many physical processes
tunneling of {\em two} particles occurs. For example, Semenov and
Dakhnovskii \cite{24,dos} found the bistability of the tunneling
trajectories in a two-proton transfer. Later, Benderskii and
co-workers extended their investigations to different
two-dimensional potentials \cite{6}. Some features of the
two-dimensional tunneling dynamics were studied in interacting
Josephson junctions \cite{11}.

Another example of two-particle tunneling is a two-proton transfer
in porphyrin systems \cite{23,dos,24}. Porphyrins are important
molecules in Biology \cite{fr} and a new area of Electronics --
molecular wires and devices \cite{tag}. There are experimental
data \cite{3,4,5,Mamaev} that clearly indicate that a tunneling
instability occurs at some critical temperature. Such a behavior
in the rate constant demonstrates the existence of bifurcations in
two-dimensional underbarrier trajectories. Apparently, this effect
requires a thorough investigation when the protons interact with a
thermal bath.

The effect of cleavage in the temperature dependence of the
current for two-dimensional systems of interacting Josephson
junctions was identified in Ref.~\cite{11}. However, there was no
experimental confirmation of this effect. In the present paper we
identify and study such an effect of unstable character observed
experimentally in tunneling constant as a function of temperature
for reactions with porphyrin. Also, we predict a stable cleavage
in the temperature dependence of current in the artificial
two-dimensional system, a pair of weakly coupled charged quantum
dots with single electrons tunneling to superlattice substrate,
and the associated fluctuations of quantum origin. We study the
emerging effect of chaotization near the critical point of
unstable cleavage for porphyrin.

In this paper, we continue to study bifurcation effects in two
dimensional tunneling. In particular, we consider two-proton
correlations of different types within the framework of
dissipative quantum tunneling (instanton) approach. Moreover, we
discuss a fine structure of bifurcations for systems with various
potential energy surfaces \cite{23,24,dos,6,Mamaev,Geldart}.

The layout of the paper is as follows.

In Section II, we introduce two-dimensional model potential energy
surfaces for a pair of interacting particles. In Section~III we
present experimental aspects of the model. In Section~IV, we study
the effect of temperature on the tunneling rate. In Sections~V and
VI, we calculate the rate for the parallel and antiparallel
tunneling and provide the analysis of the origin of bifurcation.
The effect of a promoting mode (an environment) is studied in
Section~VII.

\section{\label{Sec2}Two-dimensional potential energy surfaces}

Consider two charges that tunnel in two independent double-well
potentials $U(q_1)$ and $U(q_2)$ presented as follows
\cite{12,23,24,dos}:
\begin{eqnarray}
\label{1}
\tilde{U}(q_i) = \frac{1}{2}{\omega^2(q_i+a)^2}\theta(-q_i)
\nonumber \\
 + \left[-\Delta I +
 \frac{1}{2}{\omega^2(q_i-b)^2}\right]\theta(q_i),
 \quad i=1,2,
\end{eqnarray}
where the sum $a+b$ determines the length of a "link" in the
corresponding macrocluster fragment; $\Delta I =
\omega^2(b^2-a^2)/2$ is a bias (an asymmetry parameter of the
potential); $\theta(q_i)$ is a step function, and $\omega$ is the
frequency (see discussion in  Ref.~\cite{23}). The mass is
absorbed into the definition of $q$.

The interaction between two charges, e.g., protons, is considered
in a dipole-dipole approximation \cite{24}
\be\label{2}
V_{\mathrm{int}}(q_1,q_2) = - \frac{\alpha}{2}(q_1-q_2)^2,
\ee
where $\alpha$ is a positive constant. We use the same interaction
potential as in Ref.~\cite{11}.

\begin{figure}[tbp!]
\begin{center}
\includegraphics[width=0.6\textwidth]{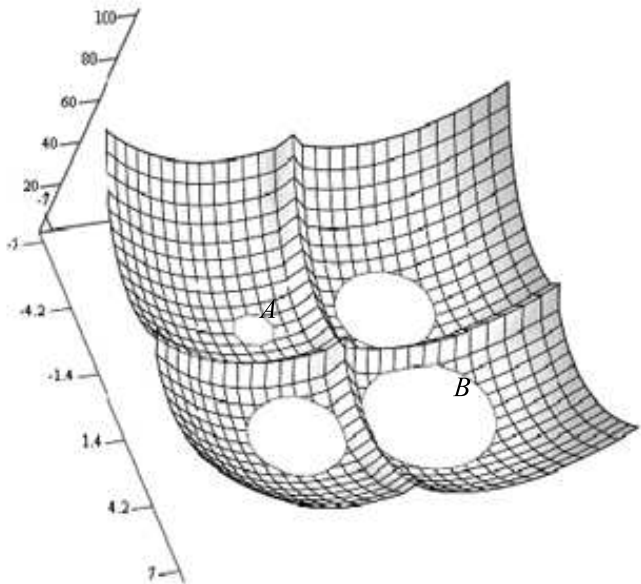}
\end{center}
\caption{ \label{Fig1} Asymmetric potential energy surface
(\ref{3}) for  parallel tunneling; $a=2$, $b=2.5$, $\alpha^*
=0.0001$. $A$ and $B$ indicate the initial and final states of the
particles, respectively. The minimum of the potential at $B$ is
lower than that of at $A$. The other two (intermediate) minima are
lower than those of at $A$ and higher than at $B$.}
\end{figure}

Thus, the total two-dimensional potential energy surface for
parallel tunneling normalized by $\omega^2$, is given by
\begin{eqnarray} U_{\mathrm{p}}(q_1,q_2) =
\frac{2\tilde{U}_{\mathrm{p}}(q_1,q_2)}{\omega^2} =
 (q_1+a)^2\theta(-q_1)
 \nonumber\\
 + \left[-(b^2-a^2)+(q_1-b)^2\right]\theta(q_1)
 + (q_2+a)^2\theta(-q_2)
 \nonumber\\
 + \left[-(b^2-a^2) + (q_2-b)^2\right]\theta(q_2)
 - \frac{\alpha^*}{2}(q_1-q_2)^2.
 \label{3}
\end{eqnarray}
Here, $\alpha^* = {2\alpha}/{\omega^2}$ is the dimensionless
parameter, $\alpha^* <1$, $\alpha \simeq e^2/(\varepsilon R_0^3)$,
$e$ is the electron charge, $R_0$ is the separation distance
between the reaction coordinates $q_1$ and $q_2$ of the tunneling
particles; $\varepsilon$ is the dielectric constant. The form of
potential energy surface (\ref{3}) is shown in Fig.~\ref{Fig1}.

\begin{figure}[tbp!]
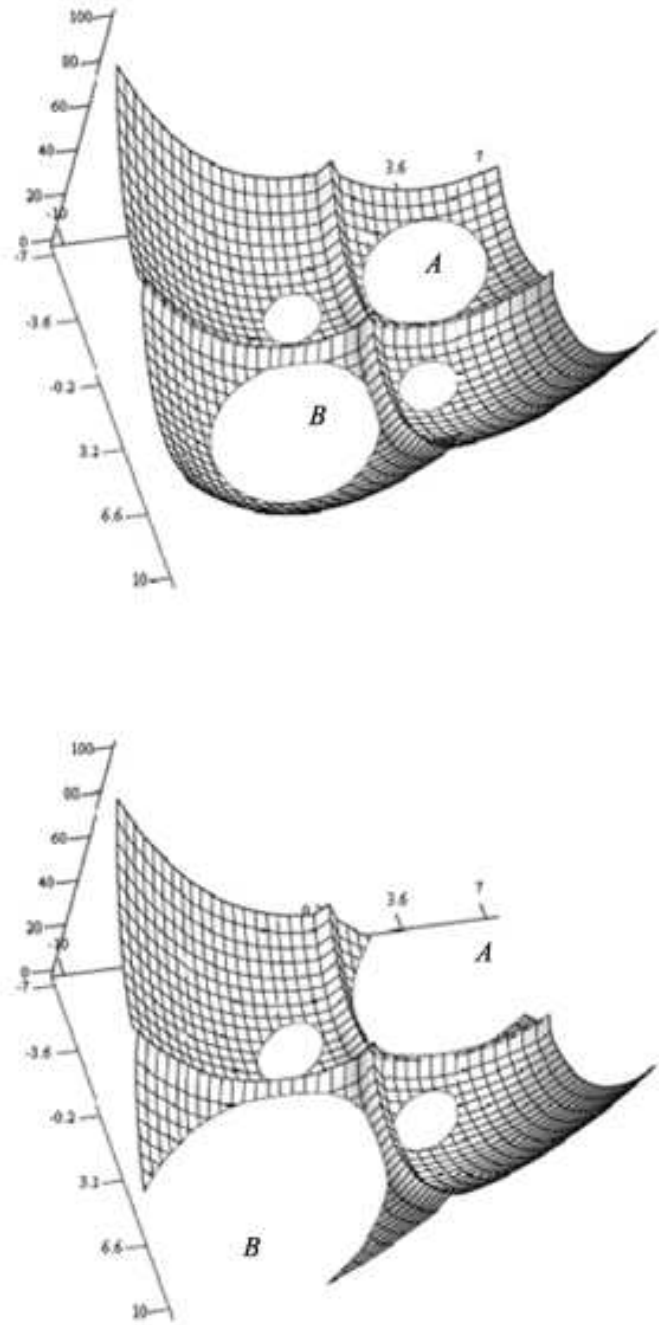

\begin{center}
\includegraphics[width=0.6\textwidth]{f2a}
\includegraphics[width=0.6\textwidth]{f2b}
\end{center}
\caption{ \label{Fig2} Asymmetric potential energy surface
(\ref{4}) for antiparallel tunneling; (a) $a=2$, $b=2.3$,
$\tilde{\alpha}^* =0.1$ (top panel); (b) $a=2$, $b=2.3$,
$\tilde{\alpha}^* =0.5$ (bottom panel). $A$ and $B$ indicate
initial and final states of the particles. The minimum of the
potential at $B$ is lower than that of at $A$. The other two
(intermediate) minima are higher than those of at $A$ and $B$.}
\end{figure}

For antiparallel transfer, the two-dimensional potential energy
with the interaction term can be defined as
\begin{eqnarray} U_{\mathrm{a}}(q_1,q_2) =
\frac{2\tilde U_{\mathrm{a}}(q_1,q_2)}{\omega^2} =
 (q_1+a)^2\theta(-q_1)
 \nonumber\\
 + \left[-(b^2-a^2) + (q_1-b)^2\right]\theta(q_1) +
 (q_2-a)^2\theta(q_2)
  \nonumber\\
 + \left[-(b^2-a^2) + (q_2+b)^2\right]\theta(-q_2) -
 \frac{\tilde\alpha^*}{2}(q_1-q_2)^2,
 \label{4}
\end{eqnarray}
where  $\tilde\alpha^* = {2\tilde\alpha}/{\omega^2}$ is a
dimensionless parameter, $\tilde\alpha^* <1$. The potential
(\ref{4}) is depicted in Fig.~\ref{Fig2}. The main difference
between the potential energy surfaces (\ref{3}) and (\ref{4}) is
in the initial location of the second particle, $\pm a$.

\begin{figure}[tbp!]
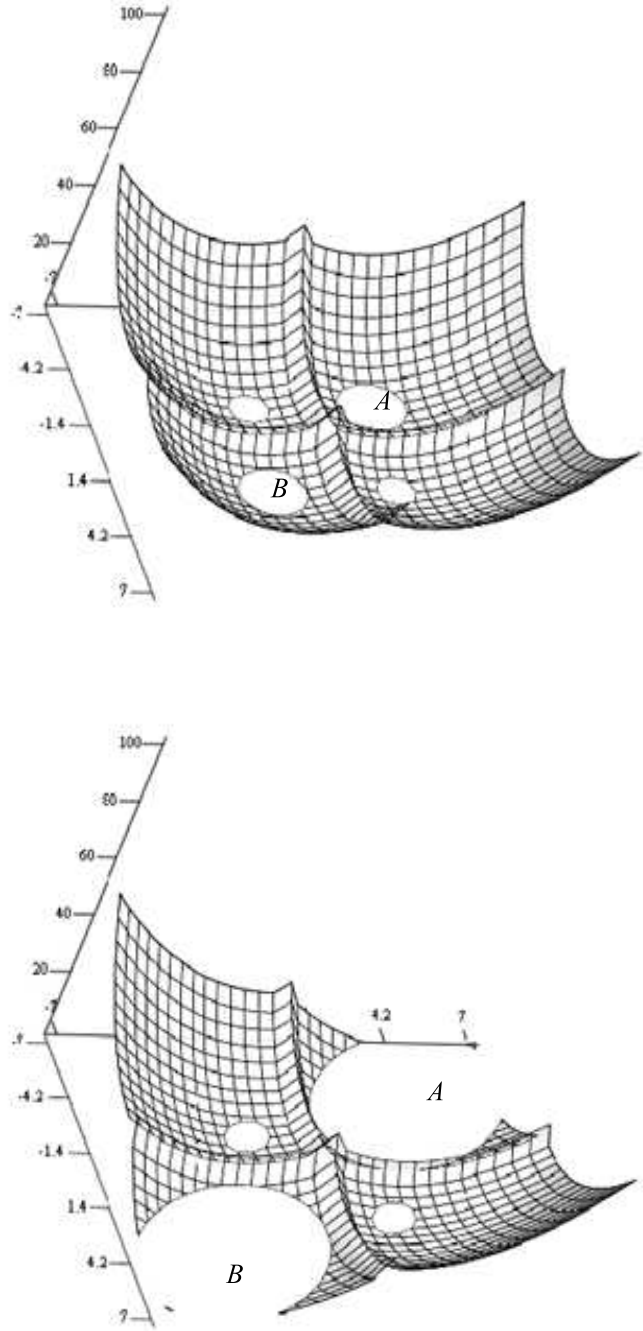

\begin{center}
\includegraphics[width=0.6\textwidth]{f3a}
\includegraphics[width=0.6\textwidth]{f3b}
\end{center}
\caption{ \label{Fig3} Symmetric potential energy surface
(\ref{5}); (a) $a=2$, $b=2$, $\tilde\alpha^{**} =0.1$ (top panel);
(b) $a=2$, $b=2$, $\tilde\alpha^{**} =0.5$ (bottom panel). $A$ and
$B$ denote the initial and final states of the particles. The
minimum of the potential energy at $B$ is equal to that of at $A$.
The other two (intermediate) minima are higher than those of at
$A$ and $B$.}
\end{figure}

The potential energy (\ref{3}) can be referred to as "parallel"
while the potential energy (\ref{4}) as "antiparallel".  We also
define a "symmetric" potential energy as a particular case of the
potential (\ref{4}) under the condition $a=b$, i.e.,
\begin{eqnarray}
U_{\mathrm{s}}(q_1,q_2) = \frac{2\tilde
U_{\mathrm{s}}(q_1,q_2)}{\omega^2} =
 (q_1+a)^2\theta(-q_1)
\nonumber\\
 + (q_1-a)^2\theta(q_1) +
 (q_2-a)^2\theta(q_2) + (q_2+a)^2\theta(-q_2)
 \nonumber\\
 - \frac{\tilde\alpha^{**}}{2}(q_1-q_2)^2,
 \label{5}
\end{eqnarray}
where the constant $\tilde\alpha^{**}<1$. The potential energy
(\ref{5}) is presented in Fig.~\ref{Fig3}.

As shown in Refs.~\cite{24, dos, Mamaev}, such model potential
energy surfaces well describe the dynamics in porphyrins (see also
Sec.~III below).

In many practical cases, the effect of a bath on particles'
tunneling should be also included. In the next section we consider
two particles embedded into a harmonic medium and linearly
interacting with the bath modes.

\section{\label{Sec3new} Experimental aspects of the model}

\begin{figure}[tbp!]
\begin{center}
\includegraphics[width=0.5\textwidth]{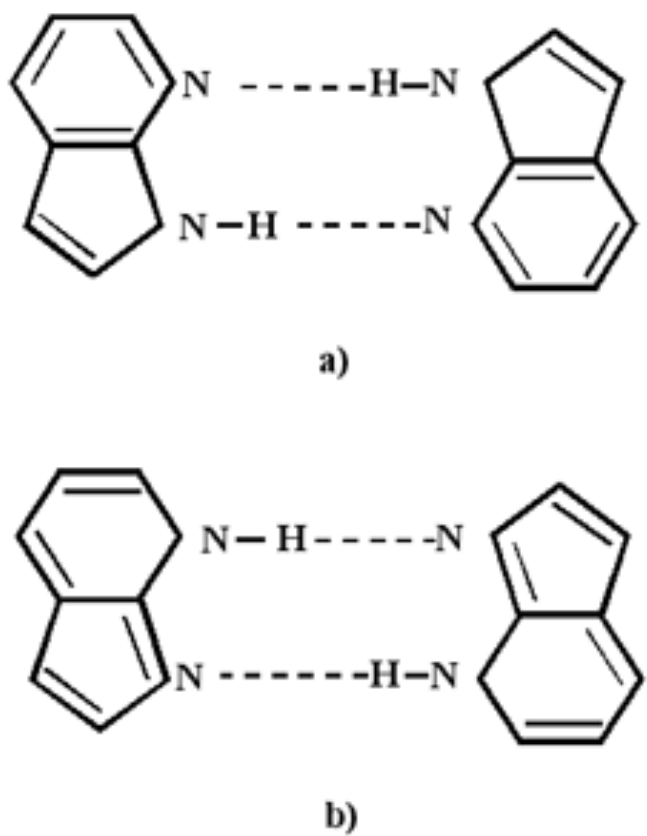}
\end{center}
\caption{\label{Fig4new} The dimer of 7-azaindole: (a) before the
proton transfer, (b) after the proton transfer.}
\end{figure}

As it was mentioned in Introduction, adiabatic tunnel dissipative
transfer appears to be very important in various physical and
chemical processes
\cite{7}-\cite{6}. It is of much interest to consider two-particle
tunnel dissipative transfer of coupled charges. For example, a
two-proton transfer can be characterized by a simultaneous
tunneling and the changing of electronic structure which forms and
breaks chemical bonds \cite{3,4,5,6}. One can consider, e.g.,
transfer of two protons in the dimer of 7-azaindole \cite{5}; see
Fig.~\ref{Fig4new}.

\begin{figure}[tbp!]
\begin{center}
\includegraphics[width=0.5\textwidth]{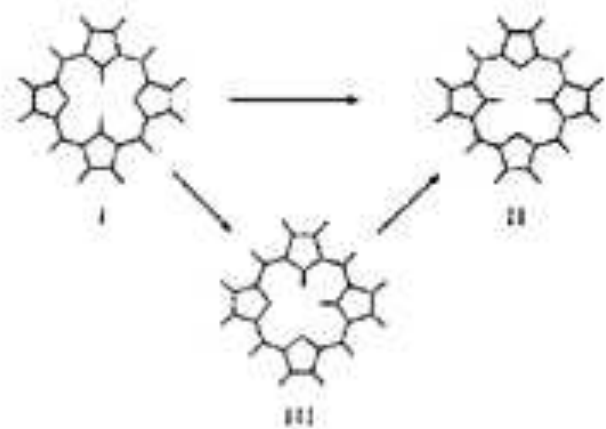}
\end{center}
\caption{\label{Fig5new} A view of two equilibrium configurations,
I and II, for the porphyrin molecule with the transfer of a pair
of hydrogen atoms between nitrogen atoms. The intermediate
(nonequilibrium) configuration III is characterized by the
hydrogen atoms residing on the neighbor nitrogen atoms.}
\end{figure}

\begin{figure}[tbp!]
\begin{center}
\includegraphics[width=0.5\textwidth]{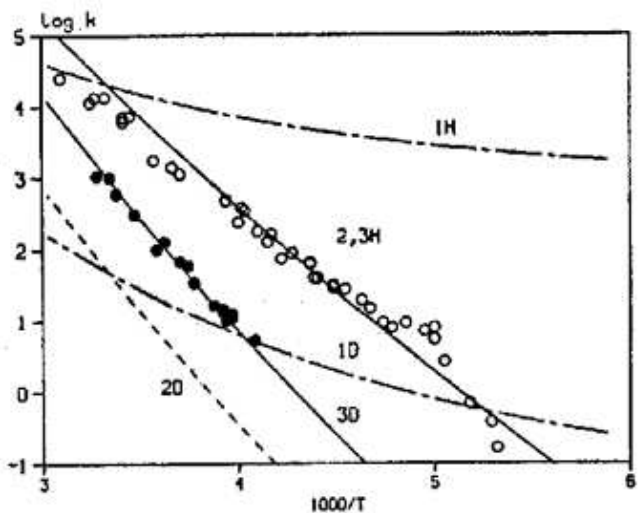}
\end{center}
\caption{\label{Fig6new} Temperature dependence of the observed
and predicted \cite{3} rate constants in the reaction of the
internal transfer of hydrogen (or deuterium) atoms in porphyrin
molecule. Open and filled circles denote experimental data for the
hydrogen and deuterium atoms respectively \cite{Hennig}. The lines
denoted by 1H and 1D represent predicted curves \cite{3} for a
synchronous transfer of weakly coupled hydrogen and deuterium
atoms respectively. The lines denoted 2H and 2D represent
predicted curves for asynchronous transfer of weakly coupled
hydrogen and deuterium atoms respectively. All the coupling
constants $J$ were scaled to reproduce the results for hydrogen
atom transfer at $T=300$ K. The used parameters are: the
stretching and bending frequencies for the hydrogen atom are 3318
and 2482 cm$^{-1}$, for the deuterium atoms are 1223 and 1094
cm$^{-1}$ respectively \cite{Gladkov}; H and D stretching
anharmonics and reduced masses are $-$80 and $-50$ cm$^{-1}$, and
1 and 2 respectively; the skeletal frequency is 380 cm$^{-1}$ and
the reduced mass is 2.8. The equilibrium separation N$-$H (N$-$D)
is 400 pm. Endothermic value for an asynchronous mechanism is
$\Delta E=-0.375$ eV, the coupling constant for synchronous
transfer is $J=0.016$ eV, the coupling constant for asynchronous
transfer of weakly coupled atoms is $J=0.23$ eV, the coupling
constant for asynchronous transfer of strongly coupled atoms is
$J=0.70$ eV.}
\end{figure}

It is particularly interesting to consider two-dimensional tunnel
dynamics of coupled protons in a symmetric case of porphyrin
molecules. For this symmetric reaction an unstable interchanging
of the regimes of tunneling (synchronous regime is replaced by
asynchronous one as temperature decreases) was observed for the
first time. The interchanging corresponds to the instability
cleavage in the dependence of the rate constant on temperature. In
some aspect this picture resembles the effect predicted for
interacting Josephson junctions \cite{11}. In Fig.~\ref{Fig5new}
we schematically present the above mentioned reaction of the
coupled protons transfer in porphyrin molecule. Fig.~\ref{Fig6new}
\cite{3} represents the experimental and the theoretically
estimated temperature dependencies of the reaction rate for the
transfer of internal hydrogen (or deuteron) atoms in porphyrin.

As it was mentioned by W.~Siebrand (Div. of Chemistry, Nat.
Research Council of Canada, Ottawa), a modified and extended
microscopic model could be used to describe transfer charge
transfer states in molecular dimers or crystals. Such states can
be represented in the form $A^+A^- \leftrightarrow A^-A^+$ that
formally corresponds to a two-electron transfer. Two types of
dimers corresponding to $+$ and $-$ combinations are separated by
the band which is associated to the frequency of the transfer.
None of these states have a constant dipole moment but an external
field can mix them and induce a quite considerable dipole moment.

In experiments, the dependence of the absorption spectrum of such
systems on the external field was measured. The results were
interpreted in terms of charge-transfer excitons. This
interpretation is valid only if the external field is sufficiently
strong, namely, to be able to break the symmetry.

Therefore, it is of much interest to investigate splitting in the
spectrum of charge-transfer dimer. In early studies one neglected
an electron-electron interaction that appears to be of much
importance here. The splitting seems to be much smaller than it
was estimated earlier. In terms of the model presented below,
electrons are attracted by Coulomb force of positive molecular
ions and repel each other. Since the instantaneous situation when
the electrons are at equal distance from the two molecules is
energetically not favorable, a high barrier for the transfer is
formed. An asynchronous transfer gives an intermediate state
corresponding to highly excited vibronic state so that this type
of transfer is not favorable due to Frank-Condon factors.

Also, this state rapidly undergoes vibronic relaxation that
corresponds to a non-radiative decay of charge-transfer states
\cite{Sebastian}. As the result, it is still important to describe
the observed properties in terms of controlled processes of
electron transfer. Although W.~Siebrand expressed some doubts on
the possibility of a direct use of the model presented below to
the two-proton transfer in porphyrin-type systems, he pointed out
that there is a possibility to use such a nonlinear two-particle
model in studying experimentally observed effectively
two-dimensional (and two-particle) electron transfer.

\begin{figure}[tbp!]
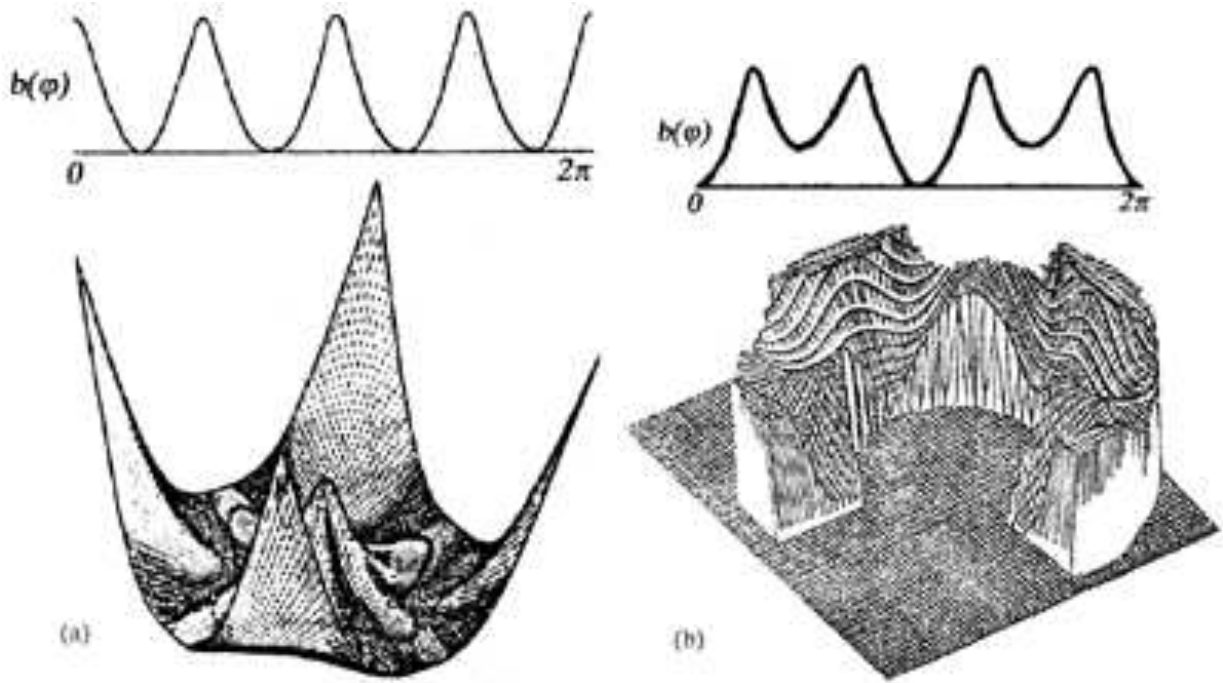

\begin{center}
\includegraphics[width=0.45\textwidth]{f7anew}
\includegraphics[width=0.45\textwidth]{f7bnew}
\end{center}
\caption{\label{Fig7new} Two-dimensional adiabatic potential and
the path characterized by a minimal energy for (a) synchronous and
(b) asynchronous mechanisms of NH-tautometry. }
\end{figure}

The microscopic model suggested in this paper allows one to
interpret the experimental data, and provides complete qualitative
and good quantitative description. Also, the model allows one to
determine the critical point in an analytically exact manner (in
the quasiclassical approximation), both  the temperature and the
coupling coefficient, that is an interesting theoretical result
which is analogous to that of Ref.~\cite{11}.  As an indirect
confirmation of applicability of the microscopic model we mention
a comparison with the results of modelling of the potential energy
surface in porphyrins \cite{6}.

For the case of 7-azaindole dimer, the protons are initially
reside closely to nitrogen atoms in the corresponding 5-chain
"rings" of the dimer. When two coupled molecules of 7-azaindole
form a dimer there may occur a cooperative (tunnel) two-proton
transfer. The protons are moved to nitrogen atoms of 6-chain
"rings" of the dimer. The proton transfer is accompanied by
changing of the structure of orbitals of electrons. A synchronous
transfer of two protons appears to be energetically preferable
while the independent tunneling of the protons becomes forbidden
due to the changing of the structure of orbitals.

In Sections below, we consider a synchronous tunnel transfer of
two weakly coupled particles. For two uncoupled particles, each of
the particles moves independently in the proper double-well
potential. We will study the effect of coupling to oscillators (an
environment) at independent ({\it asynchronous}) and cooperative
({\it synchronous}) tunneling of the particles. Also, we will
investigate fine structure of the arising two-dimensional
bifurcations which occur during the changing of regimes.

As it was mentioned above, the study of temperature-dependent NMR
spectra of porphyrins (compounds of this class are known to play
an important role in biological processes) revealed intermolecular
regrouping (see Fig.~\ref{Fig5new}). To describe this regrouping
two mechanisms are suggested: (i) the {\it synchronous} one (which
is associated to a synchronous transfer of two NH hydrogens) and
(ii) the {\it asynchronous} one (with consecutive transfer of one
of the particles).

Two-dimensional potential energy surfaces (PES) $V(r,\varphi)$
have been calculated by using INDO and we represent them in the
following analytical way:
\be
V(r,\varphi) = C [a(\varphi)(r-r_{min}(\varphi))^2 + b(\varphi)].
\ee
The synchronous case:
\begin{eqnarray}
b(\varphi) = 1.355+1.681\cos(4\varphi)+0.488\cos(8\varphi)\nonumber\\
+0.289\cos(12\varphi)+0.147\cos(16\varphi)+0.049\cos(20\varphi)\nonumber\\
+0.023\cos(24\varphi) \ \mbox{eV},
\end{eqnarray}
\be
r_{min}(\varphi) = 1.00-0.14\cos(4\varphi) \ \AA,
\ee
\be
a(\varphi) = 38.1\cos(4\varphi)+50.8 \ \mbox{eV}/\AA^2.
\ee
The asynchronous case:
\begin{eqnarray}
b(\varphi) = 0.829-0.183\cos(2\varphi)+0.821\cos(4\varphi) \nonumber \\
+0.260\cos(8\varphi)-0.077\cos(12\varphi) \ \mbox{eV},
\end{eqnarray}
\be
r_{min}(\varphi) = 1.00-0.14\cos(4\varphi) \ \AA,
\ee
\be
a(\varphi) = 19.05\cos(4\varphi)+25.4 \ \mbox{eV}/\AA^2.
\ee
The potential along the path characterized by a minimal energy and
PES $V(r,\varphi)$ are illustrated in Fig.~\ref{Fig7new}.

By comparing of the chosen model potential (see Fig.~\ref{Fig3})
with that of Fig.~\ref{Fig7new} one can see that the chosen model
potential corresponds to a realistic PES for porphyrin. More
details on PES for porphyrin have been discussed in Ref.~\cite{6}.

\section{\label{Sec3} Two-particle transition probability}

We assume that two particles independently interact with a
harmonic bath. Such interaction is considered in a bilinear
approximation. The dynamics of the environment is described by the
oscillator Hamiltonian (we use $\hbar=1$, $k_B=1$ units  with the
oscillators'  masses  equal to one),
\be\label{6}
H_{\mathrm{ph}} = \frac{1}{2}\sum\limits_{i}\left(P_i^2 +
\omega_i^2Q_i^2\right).
\ee
Each of the tunneling particles (electrons or effective charges)
interacts with the oscillator bath in the following way:
\be\label{7}
 V_{\scriptsize\textrm{p-ph}}^{(1)}(q_1, Q_i) =
q_1\sum\limits_{i}C_iQ_i,\
 V_{\scriptsize\textrm{p-ph}}^{(2)}(q_2,
Q_i) = q_2\sum\limits_{i}C_iQ_i,
\ee
As in Ref.~\cite{24}, we are interested in the transition
probability per unit time or, more precisely, only in its
exponential part which can be written in the Langer's form,
\be\label{8}
\Gamma = 2T\, \frac{\mathrm{Im}Z}{\mathrm{Re}Z}.
\ee
In such a consideration, metastable levels can be presented as
\be\label{9}
\Gamma = -2\, \mathrm{Im} E, \quad E = E_0 - i\Gamma/2.
\ee
It should be emphasized that Eq.~(\ref{8}) is valid only for
temperatures below the crossover temperature while for higher
temperatures a different prefactor is required \cite{Ankerhold}.

Equation (\ref{8}) is obtained by generalizing the expression
(\ref{9}) to nonzero temperatures
\cite{7,8,9,10,11,12,13,14,15,16,17,18,19,20,21,22}
\be\label{10}
\Gamma
 =\frac{2\sum e^{-E_{0i}/T}\mathrm{Im}E_i}{e^{-E_{0i}/T}}
 =\frac{2T\, \mathrm{Im}\sum e^{-E_{i}/T}}
 {\mathrm{Re}\sum e^{-E_{i}/T}}
 = \frac{2T\, \mathrm{Im}Z}{\mathrm{Re} Z}.
\ee
Here, $i$ labels the energy levels in the metastable state, $Z$ is
the partition function of the system, and $T$ is the temperature.

To calculate $\Gamma$, it is convenient to present $Z$ in the form
of a path integral
\cite{7,8,9,10,11,12,13,14,15,16,17,18,19,20,21,22}
\be\label{11}
Z = \prod\limits_i \int\! Dq_1Dq_2DQ_i\exp[-S\{q_1,q_2,Q_i\}].
\ee
Here, $S$ denotes an underbarrier action for the total system. The
imaginary part $\mathrm{Im}Z$ is due to the decay of the energy
states in the initial well. The validity of this approximation
requires dissipation to be strong enough so that only an
incoherent decay occurs \cite{Ankerhold}. The imaginary part in
the partition function with a double well potential energy can be
also explained due to the strong dissipation to the bath. Indeed,
the particles do not come back to their initial state. Coherent
oscillations can happen only if the interaction with bosons is
weak enough \cite{13} or the bath is in nonequilibrium state
\cite{d1,deb}.

The integral (\ref{11}) can be performed over phonon coordinates
\cite{24}  resulting in
\begin{eqnarray}
S\{q_1,q_2\} = \int\limits_{-\beta/2}^{\beta/2} \!\!\! d\tau
\biggl[\frac{1}{2}\dot q_1^2 + \frac{1}{2}\dot q_2^2 + V(q_1,q_2)
 \nonumber\\
  + \int\limits_{-\beta/2}^{\beta/2} \!\!\! d\tau'
D(\tau-\tau')[q_1(\tau)+q_2(\tau)][q_1(\tau')+q_2(\tau')])
\biggr],
\label{12}
\end{eqnarray}
where
\be\label{13}
D(\tau) =
\frac{1}{\beta}\sum\limits_{n=-\infty}^{\infty}D(\nu_n)e^{i\nu_n
\tau},
\ee
$\beta = \hbar/(k_BT)$ is  an inverse temperature (below we assume
that $\hbar=1$ and $k_B=1$), $\nu_n = 2\pi n/\beta$ is the
Matsubara's frequency, and
\be\label{14}
D(\nu_n) = - \sum\limits_i\frac{C_i^2}{\omega_i^2 + \nu_n^2}.
\ee

A trajectory that minimizes the Euclidean action $S$ can be found
from the equations of motion. In particular, we embark on the
antiparallel tunneling
\begin{eqnarray}
-\ddot q_1 + \Omega_0^2q_1 + \tilde\alpha_1q_2 +
\!\!\!\!\int\limits_{-\beta/2}^{\beta/2}\!\!\!\!d\tau'
K(\tau-\tau')[q_1(\tau')+q_2(\tau')]
 \nonumber\\
 + \omega^2a\theta(-q_1)-\omega^2b\theta(q_1) = 0,\quad
\label{15}
\end{eqnarray}
\begin{eqnarray}
-\ddot q_2 + \Omega_0^2q_2 + \tilde\alpha_1q_2 +
\!\!\!\!\int\limits_{-\beta/2}^{\beta/2}\!\!\!\!d\tau'
K(\tau-\tau')[q_1(\tau')+q_2(\tau')]
 \nonumber\\
 - \omega^2a\theta(q_2)+\omega^2b\theta(-q_2) = 0. \quad
 \label{16}
\end{eqnarray}
In Eqs. (\ref{15}) and (\ref{16}) the kernel $K$ is defined by
\be\label{17}
K(\tau) =
\frac{1}{\beta}\sum\limits_{n=-\infty}^{\infty}\xi_ne^{i\nu_n\tau}.
\ee
Here, $\xi_n$ is the Green's function defined by Eq.~(\ref{14})
without the zero-frequency term,
\be\label{18}
D(\nu_n) = -\sum\limits_{i}\frac{C_i^2}{\omega_i^2} + \xi_n.
\ee
Thus, we seek for solutions of Eqs. (\ref{15}) and (\ref{16}) by
expanding the trajectories $q_1(t)$ and $q_2(t)$ into Fourier
series,
\be\label{19}
 q_1 =
\frac{1}{\beta}\sum\limits_{n=-\infty}^{\infty}q_n^{(1)}
e^{i\nu_n\tau},
\quad
 q_2 =
\frac{1}{\beta}\sum\limits_{n=-\infty}^{\infty}q_n^{(2)}
e^{i\nu_n\tau}.
\ee
Introducing the renormalized frequency and interaction constant,
\be\label{20}
\Omega_0^2 = \omega^2 - \sum\limits_{i}\frac{C_i^2}{\omega_i^2} -
\tilde\alpha,
 \quad
 \tilde\alpha_1 = \tilde\alpha -
 \sum\limits_{i}\frac{C_i^2}{\omega_i^2},
\ee
respectively, and substituting Eqs. (\ref{19}) into
Eqs.~(\ref{15}) and (\ref{16}), we obtain that for $n=0$,
\begin{eqnarray}
 q_0^{(1)} + q_0^{(2)} =
\frac{2\omega^2(a+b)\varepsilon}{\Omega_0^2+ \tilde\alpha_1},
\nonumber\\
 q_0^{(1)} - q_0^{(2)} =
-\frac{2\omega^2a\beta}{\Omega_0^2 - \tilde\alpha_1} +
\frac{4\omega^2(a+b)\tau_0}{\Omega_0^2 - \tilde\alpha_1},
\label{21}
\end{eqnarray}
and for $n\not=0$,
\begin{eqnarray}
 q_n^{(1)} + q_n^{(2)} =
\frac{2\omega^2(a+b)(\sin\nu_n\tau_1-\sin\nu_n\tau_2)}
{\nu_n(\nu_n^2
+\Omega_0^2 + \tilde\alpha_1 +2\xi_n)}, \nonumber\\
  q_n^{(1)} - q_n^{(2)} =
\frac{2\omega^2(a+b)(\sin\nu_n\tau_1+\sin\nu_n\tau_2)}
{\nu_n(\nu_n^2
+\Omega_0^2 - \tilde\alpha_1)}.
\label{22}
\end{eqnarray}
Here, we have introduced the following notation:
\be\label{23}
\varepsilon = \tau_1-\tau_2,
 \quad
 \tau_0 = (\tau_1+\tau_2)/2.
\ee
The time instants $\tau_1$ and $\tau_2$, at which the particles
pass the top points of the barrier, are determined from the
following equations:
\be\label{24}
q_1(\tau_1)=0,
 \quad
q_2(\tau_2)=0.
\ee
Equations (\ref{24}) allow us to change the argument of the
$\theta$-function. Namely, instead of the $q_1$- and
$q_2$-dependencies, we can use a time-dependent $\theta$-function.
This reduces Eqs.~(\ref{15}) and (\ref{16}) to a linear form.

Finally, substituting the trajectory determined from
Eqs.~(\ref{19}), (\ref{21}) and (\ref{22}) into Eq.~(\ref{12}), we
arrive at the following expression for the instanton action:
\begin{eqnarray}
S = \frac{4\omega^4a(a+b)\tau_0}{\Omega_0^2-\tilde\alpha_1}
 -
 \frac{\omega^4(a+b)^2\varepsilon^2}
 {\beta(\Omega_0^2+\tilde\alpha_1)}
 -
 \frac{4\omega^4(a+b)^2\tau_0^2}
 {\beta(\Omega_0^2-\tilde\alpha_1)}
\nonumber\\
 -
 \frac{8\omega^4(a+b)^2}{\beta}
 \sum\limits_{n=1}^{\infty}
 \left[
 \frac{\sin^2\nu_n\tau_0\cdot \cos^2(\nu_n\varepsilon/2)}
      {(\nu_n^2 + \Omega_0^2 - \tilde\alpha_1)\nu_n^2}\right.
 \nonumber\\
 \left.
 +\frac{\cos^2\nu_n\tau_0\cdot \sin^2(\nu_n\varepsilon/2)}
      {(\nu_n^2 + \Omega_0^2 + \tilde\alpha_1 + 2\xi_n)\nu_n^2}
 \right].\
 \label{25}
\end{eqnarray}

\section{\label{Sec4} Parallel particle tunneling}

For the case of parallel particle tunneling, the Euclidean action
$S$ can be determined similarly to antiparallel tunneling [see
Eqs. (\ref{15}) and (\ref{16})]. The trajectory minimizing the
Euclidean action (instanton), can be determined from the equations
of motion. As in the previous section, we seek for solutions of
these equations in the form of the Fourier expansion (\ref{19}).
The time instants $\tau_1$ and $\tau_2$ are determined by
Eqs.~(\ref{24}).

In the case of parallel tunneling particles [the potential energy
(\ref{3})], the resulting Euclidean action is given as follows:
\begin{eqnarray}
 S = 2a(a+b)(\tau_1+\tau_2)\omega^2
 -
 \frac{1}{\beta}\omega^2(a+b)^2(\tau_1+\tau_2)^2
 \nonumber\\
 -
 \frac{\omega^4(a+b)^2(\tau_1-\tau_2)^2}{(\omega^2-2\alpha)\beta}
 \nonumber\\
 -
 \frac{2\omega^4(a+b)^2}{\beta}
 \sum\limits_{n=1}^{\infty}
 \biggl[
 \frac{(\sin\nu_n\tau_1+\sin\nu_n\tau_2)^2}
 {\nu_n^2(\nu_n^2+\omega^2+\xi_n)}
  \nonumber\\
 +
 \frac{(\sin\nu_n\tau_1-\sin\nu_n\tau_2)^2}
 {\nu_n^2(\nu_n^2+\omega^2-2\alpha)}
 \biggr],
\label{26}
\end{eqnarray}
where $\xi_n$ is defined by Eq.~(\ref{18}).

Below, we use the following notation:
 $$\varepsilon = \varepsilon^*\omega=(\tau_1-\tau_2)\omega,$$
 $$\tau = 2\tau^*\omega =(\tau_1+\tau_2)\omega,$$
 $$\beta^* =\beta\omega/2,$$
 $$\alpha^*=2\alpha/\omega^2,$$
 $$b^*=b/a,$$
and assume that
 $$b \geq a.$$

In the absence of interaction with an oscillator bath, i.e., at
$\xi_n=0$, the action (\ref{26}) as a function of the parameters
$\varepsilon$ and $\tau$ yields
\begin{eqnarray}
S =
 \frac{(a+b)^2\omega}{2}
 \biggl\{
       \frac{4a\tau}{a+b}
       -\frac{\tau}{a+b}\left(1+\frac{1}{1-\alpha^*}\right)
\nonumber\\
 +
 \frac{(\tau\!\!-\!\!|\varepsilon|)\alpha^*}{1-\alpha^*}
 \!+\! \mathrm{coth}\beta^*
 \!-\! \mathrm{sinh}^{\!-\!1}\beta^*
    [
     \mathrm{cosh}(\beta^*\!\!-\!\!\tau)\mathrm{cosh}\varepsilon
\nonumber\\
+ \mathrm{cosh}(\beta^*\!\!-\!\!\tau)
     - \mathrm{cosh}(\beta^*-|\varepsilon|)
    ]
\nonumber\\
 -
 (1-\alpha^*)^{-3/2}
 \bigl[-
 \mathrm{coth}(\beta\sqrt{1-\alpha^*})
\nonumber\\
  + \mathrm{sinh}^{-1}(\beta\sqrt{1-\alpha^*})
    [
     \mathrm{cosh}((\beta^*\!\!-\!\!\tau)\sqrt{1-\alpha^*})
\nonumber\\
\times (\mathrm{cosh}(\varepsilon\sqrt{1-\alpha^*}) -1)
     +
     \mathrm{cosh}((\beta^*-|\varepsilon|)\sqrt{1-\alpha^*})
    ]
 \bigr]
\biggr\}. \quad
\label{27}
\end{eqnarray}
As soon as the trajectory is found, Eqs. (\ref{24}) can be
presented in the following form:
\begin{eqnarray}
\mathrm{sinh}\varepsilon [\,\mathrm{cosh}\tau \mathrm{coth}\beta^*
-\mathrm{sinh}\tau-\mathrm{coth}\beta^*] +
\nonumber \\
 \frac{1}{1\!-\!\alpha^*}\,
\mathrm{sinh}(\varepsilon\sqrt{1\!-\!\alpha^*})
[\mathrm{cosh}(\tau\sqrt{1\!-\!\alpha^*})\mathrm{coth}
(\beta^*\sqrt{1\!-\!\alpha^*})
\nonumber \\
-\mathrm{sinh}(\tau\sqrt{1-\alpha^*})
+ \mathrm{coth}(\beta^*\sqrt{1-\alpha^*})] =0, \nonumber\\
3-\frac{4}{1+b^*} - \frac{1}{1-\alpha^*} +
\mathrm{cosh}\varepsilon [\,\mathrm{sinh}\tau \mathrm{coth}\beta^*
-\mathrm{cosh}\tau
\nonumber \\
- 1] +\mathrm{sinh}\tau \mathrm{coth}\beta^* - \mathrm{cosh}\tau
\nonumber \\
+\frac{1}{1\!-\!\alpha^*}\,\mathrm{cosh}
(\varepsilon\sqrt{1\!-\!\alpha^*})
[\,\mathrm{sinh}(\tau\sqrt{1\!-\!\alpha^*})
\nonumber \\
\times
 \mathrm{coth}(\beta^*\sqrt{1\!-\!\alpha^*})
 - \mathrm{cosh}(\tau\sqrt{1-\alpha^*}) +1]
\nonumber \\
-\frac{1}{1-\alpha^*} [\,\mathrm{sinh}(\tau\sqrt{1-\alpha^*})\,
\mathrm{coth}(\beta^*\sqrt{1-\alpha^*})
\nonumber \\
-\mathrm{cosh}(\tau\sqrt{1-\alpha^*})]=0.\quad
\label{28}
\end{eqnarray}
Simple analytic solutions of Eqs. (\ref{28}) are obtained in the
particular case when
\begin{eqnarray}
\varepsilon = (\tau_1 - \tau_2)\omega =0, \quad \forall \beta,
\ \alpha <\omega^2/2, \nonumber \\
\tau_1 = \tau_2 = \frac{\tau}{2\omega} =
\frac{1}{2\omega}\,\mathrm{arcosh}\left[\frac{1-b^*}{1+b^*}\,
\mathrm{sinh}\frac{\beta\omega}{2}\right] +\frac{\beta}{4}.
\label{29}
\end{eqnarray}
However, a complete analysis requires numerical studies.

At sufficiently low temperatures, $\omega\beta \gg 1$, for
$1<b/a<3$, and
$$\frac{b-a}{2(b+a)} \leq
\frac{2\alpha}{\omega^2}< \alpha^*_c \equiv \frac{2(b-a)}{3b-a},
$$
we finally obtain, with the exponential accuracy,
\begin{eqnarray}
e^{-\tau\sqrt{1-\alpha^*}} \simeq [3 - \frac{4}{1+b^*} -
\frac{1}{1-\alpha^*}](1-\alpha^*)^{1/(1-\sqrt{1-\alpha^*})}
\nonumber \\
\times\left\{ 1+
(1-\alpha^*)^{1/(1-\sqrt{1-\alpha^*})}[-\frac{1}{1-\alpha^*}
+(3-\frac{4}{1+b^*} \right.
\nonumber \\
\left.
-\frac{1}{1-\alpha^*})/(1-\sqrt{1-\alpha^*})]
\right\}^{-1},
\nonumber \\
e^{-\varepsilon} \simeq [3 - \frac{4}{1+b^*} -
\frac{1}{1-\alpha^*}]e^{\tau\sqrt{1-\alpha^*}} +
\frac{1}{1-\alpha^*}. \quad
\label{30}
\end{eqnarray}

The solution (\ref{30}) is valid at
\be
\beta > \beta_c \equiv
\frac{\tau\sqrt{1-\alpha^*}}{\omega}.
\label{31}
\ee

We point out that an approximate solution can be found for large
values of the parameter $b^*$ (and small $\alpha^*$). However,
below we focus on the more important solution (\ref{30}). The
analysis indicates that there are no perturbative solutions of
Eqs.~(\ref{28}) at low temperatures and small $\varepsilon$.

For $\varepsilon =0$ [see Eq.~(\ref{29})], the action (\ref{27})
results in
\begin{eqnarray}
S_{\varepsilon=0} =
\omega(b^2-a^2)\mathrm{arcosh}\left[\frac{b^*-1}{b^*+1}\,
\mathrm{sinh}\frac{\omega\beta}{2}\right]
\nonumber \\
- \frac{1}{2}\omega^2(b^2-a^2)\beta + \omega(b+a)^2
\biggl[\mathrm{cosh}\frac{\omega\beta}{2}
\nonumber \\
-\left(1+\frac{(b^*-1)^2}{(b^*+1)^2}
\mathrm{sinh^2}\frac{\omega\beta}{2}\right)^{1/2}\biggr]
\mathrm{sinh}^{-1}\frac{\omega\beta}{2}.
\label{32}
\end{eqnarray}
The action (\ref{32}) coincides (up to a factor of two) with that
of calculated in Ref.~\cite{23}. Thus, we have calculated the
two-particle Euclidean action for the case of synchronous parallel
motion of the two interacting particles.

In the symmetric case ($b^*=1$), the action (\ref{32}) reduces to
\be
S_{\!\! \scriptsize\begin{array}{c}\varepsilon\!=\!0\\[-2pt]
a\!=\!b
\end{array}}= 4\omega^2\mathrm{tanh}\frac{\omega\beta}{4}.
\label{33}
\ee
At $b^*>1$, the character of the temperature dependence is almost
the same.

At $\varepsilon\not=0$, the corresponding action $S$ can be
obtained by substituting Eq.~(\ref{30}) into Eq.~(\ref{27}) (for
brevity, this cumbersome expression is omitted). A simple analysis
shows that $S_{\varepsilon\not=0} < S_{\varepsilon=0}$. Moreover,
it appears that the difference $\Delta S = S_{\varepsilon\not=0} -
S_{\varepsilon=0}$ has a maximum at $\omega\beta \gg 1$.

\begin{figure}[tbp!]
\begin{center}
\includegraphics[width=0.6\textwidth]{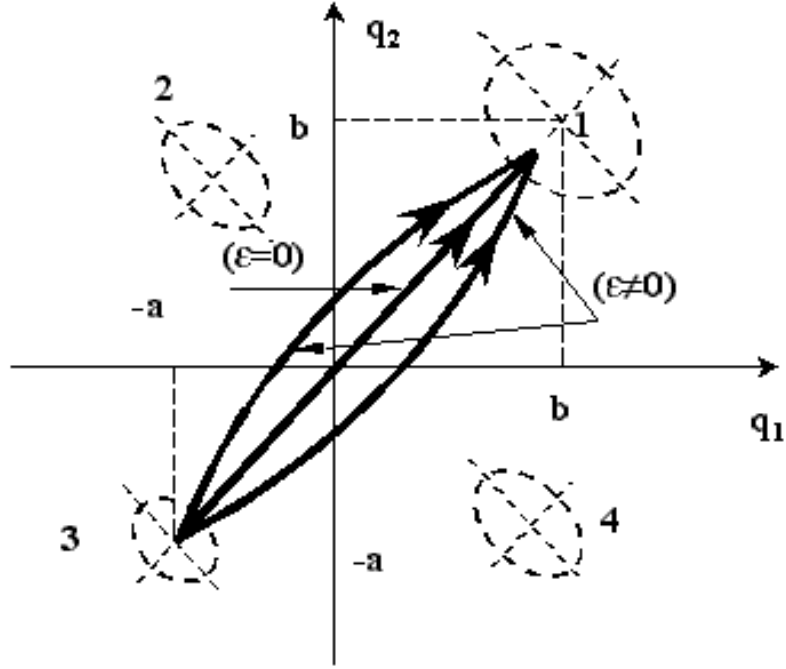}
\end{center}
\caption{\label{Fig4} Trajectories (a single path is characterized
by $\varepsilon=0$ and a splitted path is characterized by
$\varepsilon\not=0$) for two parallel moving particles, at
$\omega\beta \gg 1$; (1)-(4) label the projections of the minima
of potential energy (\ref{3}).}
\end{figure}

The tunneling paths (\ref{19}) are presented by the solutions
(\ref{29}) and (\ref{30}). At the critical point, $\beta=\beta_c$,
defined by Eq.~(\ref{31}), a relatively abrupt change in the
dynamics results in a splitting of the single trajectory
($\varepsilon=0$) into two  ($\varepsilon\not=0$), as shown in
Fig.~\ref{Fig4}.

\begin{figure}[tbp!]
\begin{center}
\includegraphics[width=0.6\textwidth]{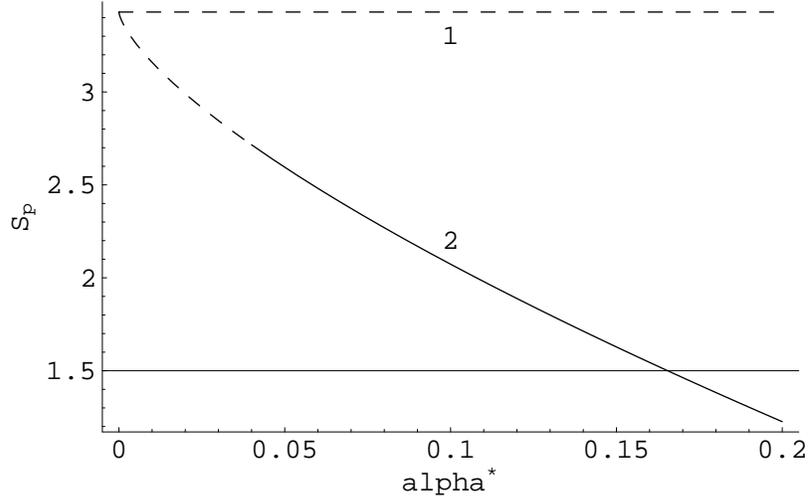}
\end{center}
\caption{ \label{Fig5} Instanton action as a function of the
interaction parameter $\alpha^*=2\alpha/\omega^2$ of the two
parallel moving particles, at $\omega\beta \gg 1 $; $S_{\mathrm
p}=S/(\omega a^2)$; (1) is the single trajectory; (2) is the split
trajectory.}
\end{figure}

At $\beta>\beta_c$, i.e., for temperatures below the critical
temperature, $T < T_c$, only a split trajectory
($\varepsilon\not=0$) occurs since $S_{\varepsilon\not=0} <
S_{\varepsilon=0}$ (see Fig.~\ref{Fig5}).

When $\beta<\beta_c$, i.e., at $T > T_c$, and $\alpha>\alpha_c$
[see Eq.~(\ref{30})], the solution is determined by a single
trajectory. We have also found that the single trajectory solution
holds in the whole temperature range for a symmetric potential
energy ($b^*=1$).

The above consideration is based on the analytic solutions
(\ref{29}) for $\tau_{1,2}$ near the critical point. A complete
analysis of Eq.~(\ref{28}) requires more detail.

\begin{figure}[tbp!]
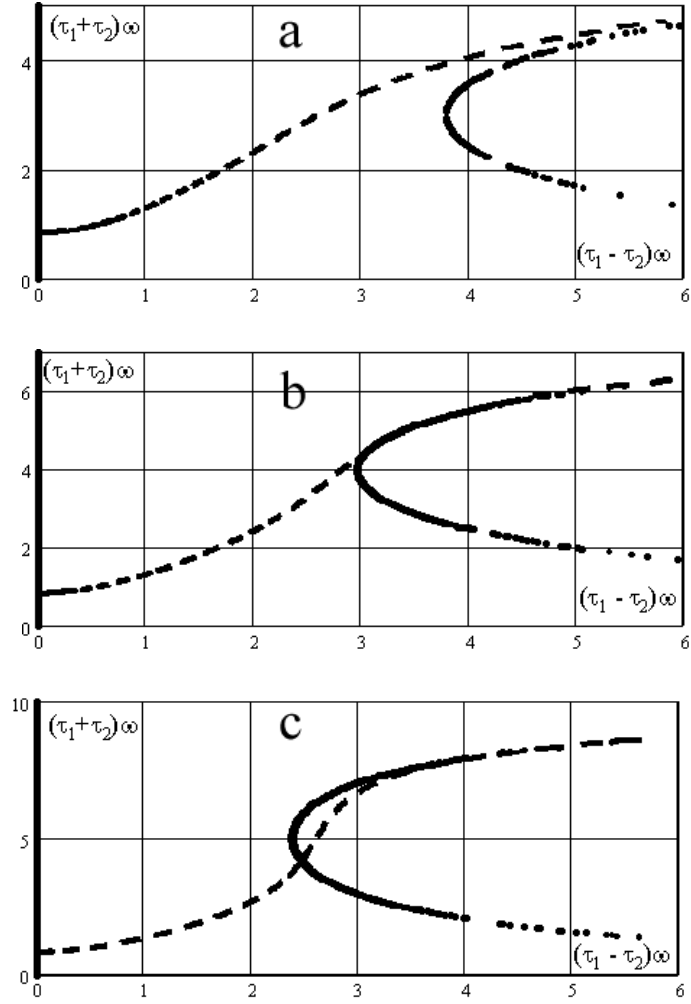

\begin{center}
\includegraphics[width=0.5\textwidth]{f6a}
 \vspace{14pt}

\includegraphics[width=0.5\textwidth]{f6b}
\vspace{14pt}

\includegraphics[width=0.52\textwidth]{f6c}
\end{center}
\caption{ \label{Fig6} Numerical solutions of transcendental
equations (\ref{28}). In addition to the studied solution
$\tau_1=\tau_2$, with larger $\beta$ there are additional
solutions, $\tau_1 \not= \tau_2$,  shown in panels (b) and (c).
Panel (b) reveals a {\em single} additional solution corresponding
to the lower value of the Euclidean action. Panel (c) shows {\rm
two} different additional solutions with the one corresponding to
the lower value of the action.}
\end{figure}

\begin{figure}[tbp!]
\begin{center}
\includegraphics[width=0.6\textwidth]{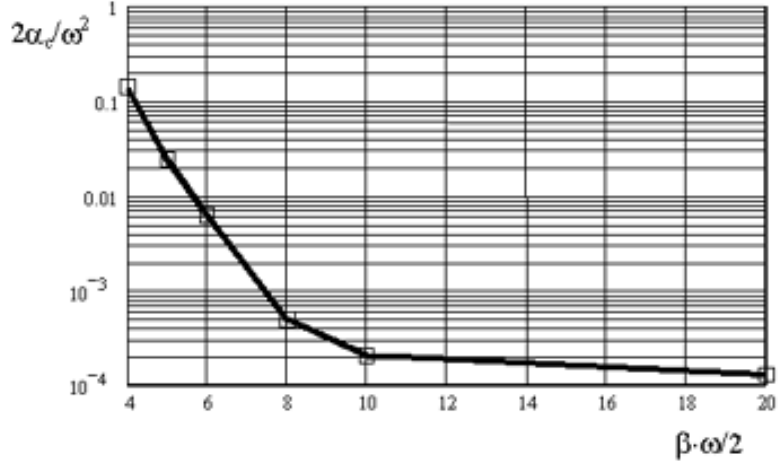}
\end{center}
\caption{ \label{Fig7} Parameter $\alpha_c$ as a function of
inverse temperature, at the fixed value of the frequency $\omega$
and asymmetry parameter $b^*$.}
\end{figure}

\begin{figure}[tbp!]
\begin{center}
\includegraphics[width=0.6\textwidth]{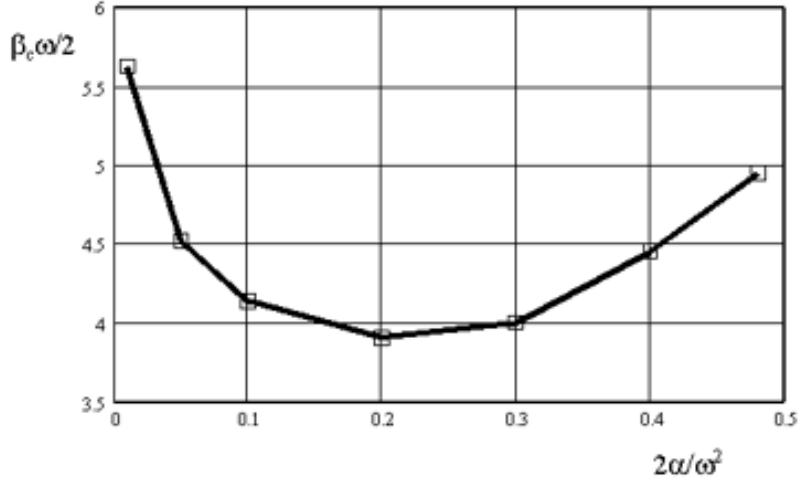}
\end{center}
\caption{ \label{Fig8} Parameter $\beta_c$ is presented as a
function of the interaction parameter of the two tunneling
particles.}
\end{figure}

The numerical study of the set of transcendental equations
(\ref{28}) reveals remarkable features of two-dimensional
tunneling. The value of the critical parameter $\alpha_c$
decreases with temperature as shown in  Figs.~\ref{Fig6},
\ref{Fig7}, and \ref{Fig8}. The temperature dependence of
$\alpha_c$ indicates to the existence of a finite low-temperature
limit. It becomes clear that the weaker the interaction between
the tunneling particle, the lower temperatures are required for
the synchronous tunneling. Since the synchronous tunneling is
valid only for $T<T_c$ [see Eq.~(\ref{31})] the dependence of
$\beta_c$ on the interaction parameter reveals that the minimum
occurs at $\alpha^*=0.2$. The nonmonotonic behavior of $\beta_c$
as a function of $\alpha^*$ can be explained by the
$\alpha^*$-dependence of $\tau$ as well.

Additionally, the curve in Fig.~\ref{Fig8} exhibits an anomalous
behavior (an increasing part of the curve) that can be explained
in the following way: a two-dimensional potential energy for
parallel tunneling is evidently deformed by an increase of the
interaction constant. Indeed, the minima of the potential surface
become lower and the distance between them is larger. Small
deformations in the potential energy surface change the
contribution produced by the temperature increase towards a
synchronous character of the tunneling. However, for sufficiently
large deformations of the potential, the situation is a quite
opposite. The significant deformations are in favor of synchronous
transfer. Thus, an increase in the interaction constant provides a
similar effect when temperature increases. Such a behavior takes
place up to the certain value of the interaction constant, beyond
which the potential energy surface becomes strongly perturbed.
Thus, Figs.~\ref{Fig7} and \ref{Fig8} are complimentary to  each
other. Consequently, Fig.~\ref{Fig8} can be viewed as a
bifurcation diagram. Indeed, the region below the curve
corresponds to the synchronous tunneling, while the region above
the curve corresponds to the asynchronous one.

Fig.~\ref{Fig6} demonstrates how the critical parameter $\beta_c$
changes from synchronous to asynchronous values with temperature.
Figs.~\ref{Fig6}b and \ref{Fig6}c reveal the existence of the
additional points, referred to as the bifurcation points,
corresponding to the change of tunneling regimes. For the two
bifurcation points shown in Fig.~\ref{Fig6}c, only one is physical
because it corresponds to the lower value of the action (the
second point is metastable). However, for sufficiently small
differences between the values of the action at these bifurcation
points, {\em both} close points contribute, thus leading to
corresponding "fluctuations" in the system during the change of
the regimes. For lower temperatures, such fluctuations become
negligible because two bifurcation points contributes differently.
Thus, a stable character of the asynchronous tunnel transfer is
achieved due to a much lower value of the action at one of the
bifurcation points.

For chemical reactions mentioned in Introduction, the effect of
the change in the tunneling regimes reveals as a {\em cleavage} in
the experimental temperature dependence of the tunneling constant.
Fine details of this instability have not been studied yet.
Nevertheless, the very existence of instability at the edge of the
bifurcation can be explained as the result of specific
fluctuations. A numerical analysis of the bifurcations in
antiparallel tunneling is given in the next section.

\section{\label{Sec5} Two-dimensional antiparallel tunneling}

For antiparallel tunneling of two particles [see the potential
energy (\ref{4})], the instanton action as a function of the
parameters $\varepsilon$ and $\tau$ is determined by
Eq.~(\ref{25}). For $\xi_n=0$, we obtain
\begin{eqnarray}
S = -\frac{\omega\tau(b^2-a^2)}{1-\tilde\alpha^*}
    -\frac{\omega(a+b)^2}{2}
    \left\{
    |\varepsilon|(1-\frac{1}{1-\tilde\alpha^*} )
    \right.
\nonumber \\
    +\frac{\mathrm{sinh}(|\varepsilon|\sqrt{1-\tilde\alpha^*})}
    {(1-\tilde\alpha^*)^{3/2}}
    - \mathrm{sinh}|\varepsilon|
    +\frac{\mathrm{cosh}(\varepsilon\sqrt{1-\tilde\alpha^*})+1}
    {(1-\tilde\alpha^*)^{3/2}}
\nonumber \\
     \times[\mathrm{sinh}(\beta^*\sqrt{1-\tilde\alpha^*})]^{-1}
     [\mathrm{cosh}((\beta^*-\tau)\sqrt{1-\tilde\alpha^*})
\nonumber \\
     -\mathrm{cosh}(\beta^*\sqrt{1-\tilde\alpha^*})]
\nonumber \\
         \left.
     +\frac{\mathrm{cosh}\varepsilon -1}{\mathrm{sinh}\beta^*}
     [\mathrm{cosh}(\beta^*-\tau) +\mathrm{cosh}\beta^*]
    \right\}.
\label{34}
\end{eqnarray}
The parameters $\varepsilon$ and $\tau$ are found from the
following set of equations [see Eq.~(\ref{24})]:
\begin{eqnarray}
 -\mathrm{sinh}\varepsilon[\mathrm{coth}\beta^*
 + \mathrm{cosh}\tau\mathrm{coth}\beta^* - \mathrm{sinh}\tau]
\nonumber \\
 + \frac{1}{1-\tilde\alpha^*}
 \mathrm{sinh}(\varepsilon\sqrt{1-\tilde\alpha^*})
   [
   \mathrm{coth}(\beta^*\sqrt{1-\tilde\alpha^*})
\nonumber \\
   -\mathrm{cosh}(\tau\sqrt{1-\tilde\alpha^*})
   \mathrm{coth}(\beta^*\sqrt{1-\tilde\alpha^*})
\nonumber \\
   +\mathrm{sinh}(\tau\sqrt{1-\tilde\alpha^*})
   ] = 0,
\nonumber \\
-1 - \frac{4}{(1+b^*)(1\!-\!\tilde\alpha^*)}
   + \frac{1}{1\!-\!\tilde\alpha^*}
\nonumber \\
   + (\mathrm{cosh}\varepsilon -1)
   (\mathrm{sinh}\tau\mathrm{coth}\beta^*
      -\mathrm{cosh}\tau)
\nonumber \\
   + \mathrm{cosh}\varepsilon
   + \frac{1}{1-\tilde\alpha^*}
     \left\{
     [\mathrm{cosh}(\varepsilon\sqrt{1-\tilde\alpha^*})+1]
     \right.
\nonumber \\
     \times[
     \mathrm{sinh}(\tau\sqrt{1-\tilde\alpha^*})
     \mathrm{coth}(\beta^*\sqrt{1-\tilde\alpha^*})
\nonumber \\
     - \mathrm{cosh}(\tau\sqrt{1-\tilde\alpha^*})
     ]
     \left.
     - \mathrm{cosh}(\varepsilon\sqrt{1-\tilde\alpha^*})
     \right\}
     =0. \quad
\label{35}
\end{eqnarray}
Simple analytic solutions of Eqs. (\ref{35}) can be obtained in
the following form:
\begin{eqnarray}
    \varepsilon = (\tau_1-\tau_2)\omega =0, \quad \forall \beta, \
    \tilde\alpha < \omega^2/2,
\nonumber \\
    \tau_1 = \tau_2 = \frac{\tau}{2\omega}
\nonumber \\
    =
    \frac{1}{2\omega\sqrt{1\!-\!\tilde\alpha^*}}\
    \mathrm{arcosh}
    \left[
    \frac{1\!-\!b^*}{1\!+\!b^*}
    \mathrm{sinh}(\frac{\beta\omega}{2}\sqrt{1\!-\!\tilde\alpha^*})
    \right]
    + \frac{\beta}{4}.
\label{36}
\end{eqnarray}

Similarly to the case of parallel tunneling, we obtain that at low
temperatures, $\omega\beta \gg 1$, with the exponential accuracy,
\begin{eqnarray}
 e^{-\tau\sqrt{1-\tilde\alpha^*}} \simeq
   \frac{A(1-\tilde\alpha^*)^{1/\gamma}}
   {
   1- (1-\tilde\alpha^*)^{1/\gamma}({A}/{\gamma}
   - ({1-\tilde\alpha^*})^{-1})
   },
   \nonumber \\
   e^{\varepsilon} \simeq A e^{\tau\sqrt{1-\tilde{\alpha}^*}}
   - \frac{1}{1-\tilde\alpha^*}.
\label{37}
\end{eqnarray}
Here,
$$
A = -1 - \frac{4}{(1+b^*)(1-\tilde\alpha^*)}
       + \frac{3}{1-\tilde\alpha^*}, \quad
       \gamma = 1- \sqrt{1-\tilde\alpha^*},
$$
while $\tilde\alpha^*$, $b^*$, $\varepsilon$, and $\tau$ are
determined as for the parallel transfer.

The solution (\ref{37}) is valid at $\tilde{\alpha}^*_{c1} <
\tilde\alpha^* < \tilde{\alpha}^*_{c2}$ , where the lower and
upper bounds $\tilde{\alpha}^*_{c1}$ and $\tilde{\alpha}^*_{c2}$
are derived from a cumbersome transcendental equation (for brevity
it is not presented here). Particularly, in the symmetric case,
$b^*=1$, we obtain the condition in a simple analytic form, $1/4 <
2\tilde{\alpha}/\omega^2 < 1$.

Furthermore, an approximate solution can be found for large values
of the parameter $b^*=b/a$ (and small $\tilde{\alpha}^*$).
However, we restrict our analysis to the more important physical
solution (\ref{37}).

The $\beta$-dependent solution (\ref{37}) is valid for $\beta >
\tilde{\beta}_c$, where
\begin{eqnarray}
\tilde{\beta}_c \equiv - \frac{1}{\omega\sqrt{1-\tilde{\alpha}^*}}
        \ln
        \frac{A(1-\tilde{\alpha}^*)^{1/\gamma}}
        {1- (1-\tilde{\alpha}^*)^{1/\gamma}(\frac{A}{\gamma} -
        \frac{1}{1-\tilde{\alpha}^*})}.
\label{38}
\end{eqnarray}

At $\omega\beta \gg 1$, the solutions of Eqs.~(\ref{35}) can be
found perturbatively (for small $\varepsilon$) with given values
of the parameters $(b-a)/(b+a)$ and $\tilde\alpha^*$. At
$\varepsilon=0$ [the solution (\ref{36})], the action (\ref{34})
yields:
\begin{eqnarray}
S =
 \frac{\omega(b^2-a^2)}{(1-\tilde\alpha^*)^{3/2}}\,
 \mathrm{arcosh}
 \left[
 \frac{b-a}{b+a}\ \mathrm{sinh}
 \frac{\omega\beta\sqrt{1-\tilde\alpha^*}}{2}
  \right]
\nonumber \\
 - \frac{\omega^2\beta(b^2-a^2)}{2(1-\tilde\alpha^2)}
 + \frac{\omega(b+a)^2}{(1-\tilde\alpha^*)^{3/2}}
 \left[
 \mathrm{coth}\frac{\omega\beta\sqrt{1-\tilde\alpha^*}}{2}
 \right.
\nonumber \\
  \left.
 -\left(
   \mathrm{sinh}^{-2}\frac{\omega\beta\sqrt{1-\tilde\alpha^*}}{2}
   + \frac{(b-a)^2}{(b+a)^2}
 \right)^{\! 1/2}
 \right].
\label{39}
\end{eqnarray}

For the symmetric potential ($a=b$) and $\varepsilon=0$, we obtain
that (see Fig.~\ref{Fig7})
\be
S = \frac{4\omega
 a^2}{(1-\tilde\alpha^*)^{3/2}}\
 \mathrm{tanh}\frac{\omega\beta\sqrt{1-\tilde\alpha^*}}{4}.
\label{40}
\ee

\begin{figure}[tbp!]
\begin{center}
\includegraphics[width=0.6\textwidth]{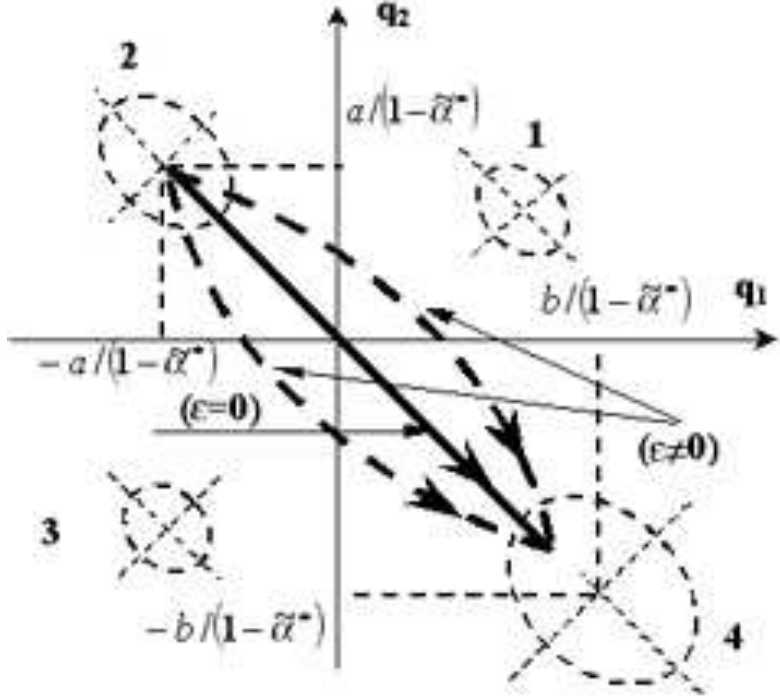}
\end{center}
\caption{ \label{Fig9} Trajectories (the basic path characterized
by $\varepsilon=0$ and the split one is characterized by
$\varepsilon\not=0$) at $\omega\beta \gg 1$ of two antiparallel
tunneling particles; (1)-(4) denote the projections of the minima
of the potential energy surface $U_{\mathrm{a}}(q_1,q_2)$ defined
by Eq.~(\ref{4}).}
\end{figure}

We do not present here a cumbersome expression for
$S_{\varepsilon\not=0}$ which one can obtain by substituting the
solutions $\tau$ and $\varepsilon$ into Eq.~(\ref{34}). A simple
analysis reveals that $S_{\varepsilon\not=0} > S_{\varepsilon=0}$
for $\beta > \tilde\beta_c$  and for relevant $\tilde\alpha^*$.
Similarly to parallel transfer, the tunnel paths can be found from
Eqs.~(\ref{36}) and (\ref{37}). These trajectories on the
$(q_1,q_2)$-plane are shown in Fig.~\ref{Fig9}.

As for parallel tunneling and at $\beta>\tilde\beta_c$,  the
pair-tunneling changes from a single to double trajectory regime.
In contrast to the parallel tunneling, such a splitting occurs at
any values of the parameters of the potential. At
$\beta>\tilde\beta_c$, we have $S_{\varepsilon\not=0} >
S_{\varepsilon=0}$ and, therefore, $S_{\varepsilon=0}$ determines
the tunneling rate. At $\beta < \tilde\beta_c$, the two
degenerated trajectories are transformed into a single trajectory,
$q_1=-q_2$, corresponding to synchronous antiparallel transfer.

For single particle tunneling, there is only a single tunneling path
(instanton) minimizing the action. Hence, there are two
different types of trajectories for the pair of interacting
particles. Namely, the main contribution to the instanton action
is determined either by the single  or by the double-degenerated
paths depending on the value of $\beta$. We also point out that in
the case of the parallel tunneling, the particles do not
simultaneously pass the top points of the barrier,
$\tau_1\not=\tau_2$. This means that the tunneling transfer is
asynchronous.

\begin{figure}[tbp!]
\vspace{14pt}
\begin{center}
\includegraphics[width=0.6\textwidth]{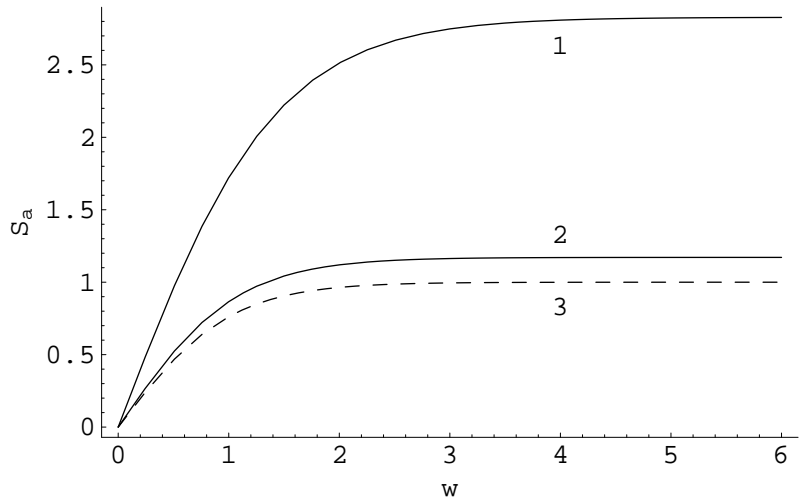}
\end{center}
\caption{ \label{Fig10} Instanton action for antiparallel
tunneling ($\varepsilon=0$, $a=b$) as a function of an inverse
temperature at the two different values of the interaction
parameter; $S_{\mathrm a}=S/(4\omega a^2)$, $w=\omega\beta/4$; (1)
$\tilde\alpha^*=0.5$; (2) $\tilde\alpha^*=0.1$; (3) a dashed line
corresponds to the action (\ref{33}) for the parallel transition
($\varepsilon=0$, $a=b$).}
\end{figure}

At small values of the interaction parameter $\alpha^*$ [see
Eq.~(\ref{30})] and at temperatures such that $\beta<\beta_c$ [see
Eq.~(\ref{31})], there is no splitting of a single path
($q_1=q_2$). Therefore, the particles pass the top  of the
barriers  at the same instants ($\tau_1=\tau_2$). Consequently,
the transfer of the particles is synchronous. The temperature
dependence for the antiparallel transfer action is plotted in
Fig.~\ref{Fig10} at various $\tilde\alpha^*$.

The type of the interaction given by Eqs.~(\ref{2})-(\ref{5}) is
such that it does not affect the motion along the "center of mass"
coordinate, $q_1=q_2$. For this reason, the Euclidean action is
independent of the interaction parameter as for parallel transfer.
Since the state of the interacting system characterized by a
maximal value of the relative coordinate, $q_1=-q_2$, is
preferable (as it provides the lower action), it becomes clear
that the instanton action decreases with the interaction parameter
in the parallel transfer along the degenerated tunnel
trajectories, and increases with the interaction parameter for the
antiparallel tunneling.

For antiparallel tunneling, synchronous transfer ($\tau_1=\tau_2$)
takes place, while asynchronous transfer is forbidden due to the
greater contribution to the Euclidean action (see
Fig.~\ref{Fig10}).

The validity condition for weakly interacting
instanton-antiinstanton pairs, in the adiabatic approximation, was
discussed in Ref.~\cite{23}.

\begin{figure}[tbp!]
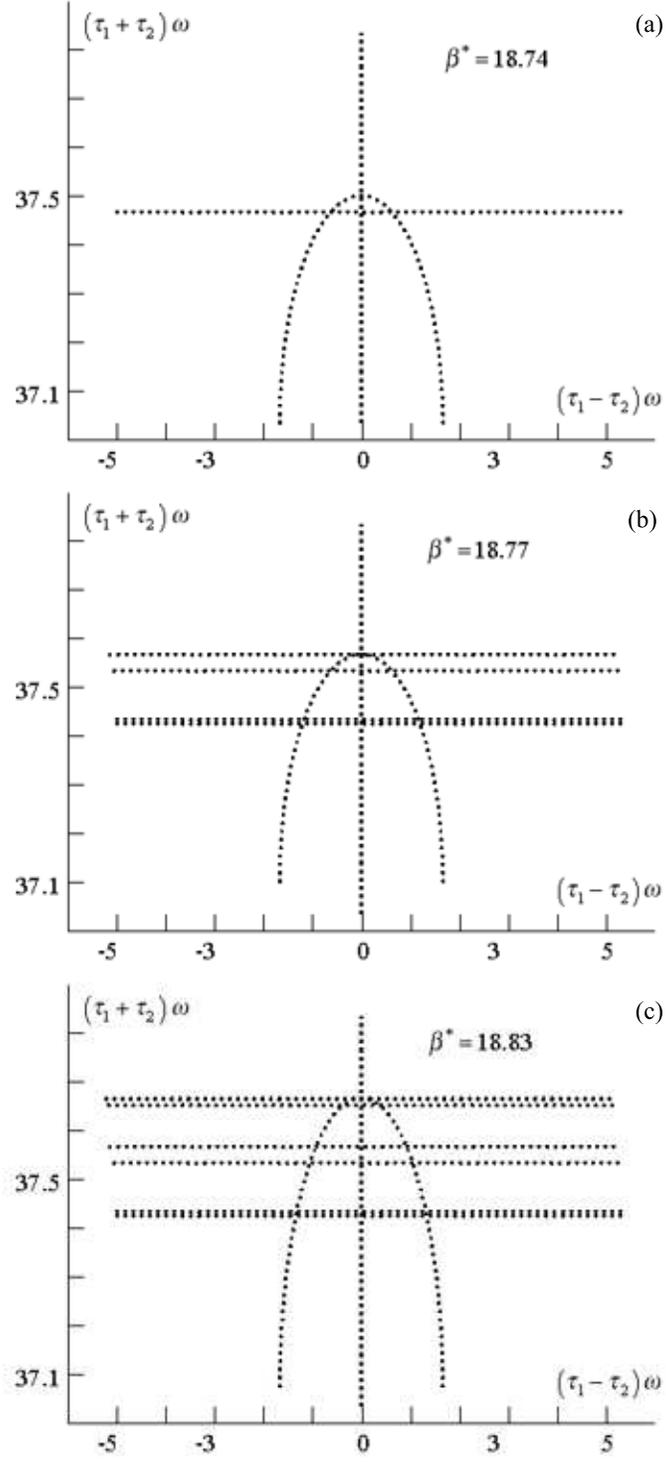

\vspace{14pt}
\begin{center}
\includegraphics[width=0.5\textwidth]{f11a}
\includegraphics[width=0.5\textwidth]{f11b}
\includegraphics[width=0.5\textwidth]{f11c}
\end{center}
\caption{ \label{Fig11} Numerical solutions of transcendental
equation (\ref{35}). In addition to the studied solution
$\tau_1=\tau_2$, with respect to $\beta$, there are the additional
solutions, $\tau_1\not=\tau_2$, shown in panels (a), (b) and (c),
that correspond to one, four, and six (pairs) additional
solutions, respectively.}
\end{figure}

The numerical analysis of transcendental equation~(\ref{35})
reveals interesting features for a transition region between the
tunneling regimes, i.e., a fine structure near the first
bifurcation point for the antiparallel transfer. The numerical
results are presented in Fig.~\ref{Fig11}. We found that, in
addition to the first bifurcation point characterized by the two
solutions (Fig.~\ref{Fig11}a), there exist additional bifurcation
points at lower temperatures, i.e., the four pairs
(Fig.~\ref{Fig11}b), the six pairs (Fig.~\ref{Fig11}c), and there
are even twelve pairs of additional solutions at $\beta^*=19.2009$
$(\alpha^*=0.05)$, etc. We refer to this phenomenon as
multiplication of bifurcations or a {\em cascade of bifurcations}.
Such an effect resembles a scenario of transition to chaos.

Although the synchronous regime is preferred due to the minimal
instanton action, in a certain temperature range its value is
comparable with those of corresponding to the cascade solutions.
As a result, {\em quantum fluctuations} of a nonregular character
occur in contrast to the parallel transfer. The antiparallel
tunneling is, thus, characterized by the instability of the
transition due to the  synchronous to asynchronous behavior. Such
instabilities are similar to continuous second-order phase
transition, while the parallel tunneling is viewed as a step
process similar to phase transition of a first order (see
Fig.~\ref{Fig6}). The dependencies shown in Figs.~\ref{Fig7} and
\ref{Fig8}, $\beta_c(\alpha)$ and $\alpha_c(\beta)$, for the
antiparallel transfer are found to be of the same character as
those of the parallel transfer.

In summary, our investigation reveals a quite complicated fine
structure of the transition for parallel and antiparallel
tunneling of two particles with different degenerate trajectories
leading to the bifurcation cascade.

\section{\label{Sec6} Effect of a promoting mode}

In this section, we study the effect of a heat bath on tunneling
transition of two interacting particles. In many tunneling
reactions,  interaction with a vibrational subsystem can often be
approximated by interaction with a single vibrational mode (a
so-called promoting mode). As it follows from Eq.~(\ref{12}), a
heat bath affects only the dynamics of the center of mass
($q_1=q_2$). Therefore, in the case of an antiparallel motion, the
medium does not affect the rate constant, while for parallel
tunneling it essentially contributes to the transfer rate. For
both parallel and antiparallel transfer, bilinear interaction of
the particles with a single oscillator can make a qualitative
change in the character of tunneling.

At small values of the interaction parameter between the two
tunneling particles  dissipative effects become important
\cite{23}. In two dimensions the dissipative effects are more
pronounced for parallel rather than antiparallel tunneling. The
latter increases with  temperature with a considerable
contribution to the prefactor. In the present work, we are
interested in the tunneling rate assuming only exponential
evolution of the transition probability. However, nonexponential
evolution can occur in an nonequilibrium environment
\cite{d1,d2,deb,dlw}. Such a case is not discussed here.
Accordingly, a reservoir is assumed to be in thermodynamic
equilibrium, i.e., the tunneling transition is rather slow
compared to the thermodynamic relaxation.  Thus, we assume that
dissipation affects the value of instanton action only.

For the case of the antiparallel tunnel transfer, the action
(\ref{25}) can be calculated with the following vibronic Green's
function:
\be
D(\nu_n) = - \frac{C^2}{\nu_n^2 + \omega_L^2},
\label{41}
\ee
where $\omega_L$ is the frequency of the vibrational mode. After some
tedious calculations, one obtains the following expression for the
instanton action (\ref{25}):
\begin{eqnarray}
S = \frac{2\omega^4(a^2-b^2)\tau_0}{\Omega_0^2}
   - \frac{2\omega^4(a+b)^2}{\beta}
\nonumber \\
   \times\left\{
   \right.
   -\frac{\beta\mathrm{sinh}(\Omega_0(\beta/2-\tau_0))
         \mathrm{sinh}(\Omega_0\tau_0)}
         {2\Omega_0^3\mathrm{sinh}(\Omega_0\beta/2)}
\nonumber \\
   + \sum\limits_{i=1}^{2}\frac{\beta(\Omega_i^2-\omega_L^2)
   \mathrm{cosh}(\Omega_i(\beta/2-\tau_0))
                 \mathrm{cosh}(\Omega_i\tau_0)}
          {(-1)^i2\Omega_i^3(\Omega_1^2-\Omega_2^2)
          \mathrm{sinh}(\Omega_i\beta/2)}
\nonumber \\
   + \frac{\beta\varepsilon}{4(\omega^2 -2C^2/\omega_L^2)}
   - \frac{\beta\varepsilon}{4\Omega_0^2}
   + \frac{\beta\mathrm{sinh}(\varepsilon\Omega_0)}{4\Omega_0^3}
\nonumber \\
   +\sum\limits_{i=1}^{2}\frac{\beta(\Omega_i^2-\omega_L^2)
   \mathrm{sinh}(\varepsilon\Omega_i)}
         {(-1)^i4\Omega_i^3(\Omega_1^2-\Omega_2^2)}
   +\frac{\beta\mathrm{cosh}(\varepsilon\Omega_0)}
         {4\Omega_0^3\mathrm{sinh}(\beta\Omega_0/2)}
\nonumber \\
    \times [
            \mathrm{cosh}(\Omega_0(\beta/2-2\tau_0))
           - \mathrm{cosh}(\Omega_0\beta/2)
           ]
\nonumber \\
   -\sum\limits_{i=1}^{2}\frac{\beta(\Omega_i^2-\omega_L^2)
                              \mathrm{cosh}(\varepsilon\Omega_i)}
    {(-1)^i4\Omega_i^3(\Omega_1^2-\Omega_2^2)
    \mathrm{sinh}(\beta\Omega_i/2)}
\nonumber \\
    \times [
            \mathrm{cosh}(\Omega_i(\beta/2-2\tau_0))
            + \mathrm{cosh}(\Omega_i\beta/2)
           ]
  \left.
     \right\}.
\label{42}
\end{eqnarray}
Here, we have introduced the following notation:
\begin{eqnarray}
 \Omega_0^2 = \omega^2 -\alpha^2, \
 \Omega_1^2
 = \frac{1}{2}(\omega^2\!+\!\omega_L^2
   \!+\!\sqrt{(\omega^2\!-\!\omega_L^2)^2+8C^2}\ ),
\nonumber \\
 \Omega_2^2
 = \frac{1}{2}(\omega^2\!+\!\omega_L^2
   \!-\!\sqrt{(\omega^2\!-\!\omega_L^2)^2+8C^2}\ ).
\nonumber
\end{eqnarray}

For the parallel tunneling transition, a corresponding action can
be found in a similar way. The action (\ref{27}) as a function of
the parameters $\varepsilon^*=\tau_1-\tau_2$ and
$\tau^*=(\tau_1+\tau_2)/2$, with the vibronic frequency $\omega_L$
and coupling constant $C$, yields (see also Ref.~\cite{23}):
\begin{eqnarray}
S \!=\!
 (a\!+\!b)(3a\!-\!b)\omega^2\tau^*\!
 -\frac{\omega^4(a\!+\!b)^2\varepsilon^*}{2(\omega^2\!-\!2\alpha)}
 -\frac{4\omega^2(a\!+\!b)^2\tau^{*2}}{\beta}
\nonumber \\
 -\frac{4\omega^4(a+b)^2\varepsilon^{*2}}{(\omega^2-2\alpha)\beta}
 -\frac{\omega^2(a+b)^2}{2\tilde\gamma}
 \sum\limits_{i=1}^{2}
 \frac{(-1)^i(\omega^2-x_{3-i})}{\sqrt{x_i}}
\nonumber \\
 \times
\left[
       \mathrm{coth}(\frac{\beta\sqrt{x_i}}{2})
       - \mathrm{sinh}^{-1}(\frac{\beta\sqrt{x_i}}{2})
\right.
         \left(
         \mathrm{cosh}[(\frac{\beta}{2}-2\tau^*)\sqrt{x_i}]
         \right.
\nonumber \\
          -\mathrm{cosh}[(\frac{\beta}{2}-\varepsilon^*)\sqrt{x_i}]
          +\frac{1}{2}\mathrm{cosh}[
          (\frac{\beta}{2}-2\tau^*-\varepsilon^*)\sqrt{x_i}]
\nonumber \\
\left.
         \left.
          +\frac{1}{2}\mathrm{cosh}[
          (\frac{\beta}{2}-2\tau^*+\varepsilon^*)\sqrt{x_i}]
         \right)
\right]
 + \frac{\omega^4(a\!+\!b)^2}{2(\omega^2\!-\!2\alpha)^{3/2}}
\nonumber \\
 \times
\left[
       -\mathrm{coth}(\frac{\beta}{2}\sqrt{\omega^2\!-\!2\alpha})
       +\mathrm{sinh}^{-1}(\frac{\beta}{2}\sqrt{\omega^2\!-\!2\alpha})
\right.
\nonumber \\
            \left(\!\!
           -\mathrm{cosh}[(\frac{\beta}{2}\!-\!2\tau^*)
           \sqrt{\omega^2\!-\!2\alpha}]
          +\mathrm{cosh}[(\frac{\beta}{2}\!-\!\varepsilon^*)
          \sqrt{\omega^2\!-\!2\alpha}]
            \right.
\nonumber \\
+\frac{1}{2}\mathrm{cosh}[
(\frac{\beta}{2}-2\tau^*-\varepsilon^*)\sqrt{\omega^2-2\alpha}]
\nonumber \\
\left.
              \left.
+\frac{1}{2}\mathrm{cosh}[
(\frac{\beta}{2}-2\tau^*+\varepsilon^*)\sqrt{\omega^2-2\alpha}]
              \right)
\right].\quad
\label{43}
\end{eqnarray}
Here, we have denoted
\begin{eqnarray}
x_{1,2} = \frac{1}{2}(\omega^2+\omega_L^2+\frac{C^2}{\omega_L^2})
          \mp \frac{1}{2}\tilde\gamma,
\nonumber \\
\tilde\gamma =
\sqrt{(\omega^2+\omega_L^2+\frac{C^2}{\omega_L^2})^2
          -4\omega^2\omega_L^2}.
\nonumber
\end{eqnarray}
For particular values of the interaction constant $\alpha$ and in
the absence of interaction with the oscillator bath, the critical
temperature $T_c$ (at which the synchronous and asynchronous
tunnel regimes interchange) is found from Eqs.~(\ref{31}) and
(\ref{38}). These equations can be generalized to non-zero
interaction with the promoting mode. Typically, the critical
temperature is found to be in the range from {\em 10 K} to {\em
400 K}. In glasses, $T_c$ can  be very small while for chemical
reactions it can be rather large. Additionally, $T_c$ depends on
the mean distance between the particles and, therefore, on their
concentration.

Quantum tunneling  is important \cite{24} when
$k_BT_c/(\hbar\omega) \leq 1$.  Therefore, the symmetry breaking
effects can take place at relatively high temperatures depending
on the ''frequency'' of a barrier. For example, for porphyrins the
critical temperature $T_c$ is estimated to be about {\em 200 K}.

\section{\label{Sec7} Conclusions}

In the single instanton approximation, we have calculated the
Euclidean action (\ref{12}) for the models characterized by the
different adiabatic potential energy surfaces, (\ref{3}),
(\ref{4}) and (\ref{5}), and made a detailed comparative analysis
of the tunneling rate for two interacting particles moving in
parallel or antiparallel within a dissipative environment. We have
also assumed exponential evolution of the transition probability
\cite{23,d1,d2,deb,dlw}.

We have shown that the change in a  tunneling regime from
synchronous to asynchronous transfer for parallel transition
occurs as a step process, similar to  phase transition of a first
order, while for the antiparallel transfer it resembles second
order phase transition.

We have explained the effect of a {\em cleavage} in the
experimentally observed \cite{3,4,5,Mamaev} temperature dependence
of the reaction rate for two tunneling particles. It has been
shown that the effect of symmetry breaking is stable for the
parallel and unstable for the antiparallel transfer, as it is
observed experimentally for some porphyrin systems
\cite{3,4,5,Mamaev}. We have found a complicated fine structure in
the bifurcation region due to the {\em quantum fluctuations} for
the parallel two-dimensional tunneling. For the antiparallel
tunneling, the contribution of four, six, twelve, etc., pairs of
trajectory becomes important resembling the transition to {\sl
chaos}.

Additionally, we have studied interaction of two particles with
phonons. Such coupling essentially modifies the antiparallel and
parallel transitions in different ways. As it follows from
Eq.~(\ref{12}), the interaction with the reservoir does not change
the dynamics of the center of mass for the antiparallel motion,
while it makes a significant contribution to the transfer rate for
the parallel transfer. Finally, Eq.~(\ref{31}) determines the
validity condition for temperatures beyond of which stable
two-dimensional synchronous tunneling
correlations of all the kinds occur.\\

The authors would like to thank A.~I.~Larkin and B.~I.~Ivlev for
the stimulating interest in this work.

\end{document}